# Three-Dimensional, Time-Dependent Simulation of Free-Electron Lasers with Planar, Helical, and Elliptical Undulators

H.P. Freund,[1] P.J.M. van der Slot,[1,2] D.L.A.G. Grimminck[3], I.D. Setija[3], and P. Falgari[4]

[1]Department of Electrical and Computer Engineering, Colorado State University, Fort Collins, Colorado, USA
[2]Mesa+ Institute for Nanotechnology, University of Twente, Enschede, the Netherlands
[3]ASML B.V., Veldhoven, the Netherlands
[4]LIME B.V., Eindhoven, the Netherlands

Free-electron lasers (FELs) have been built ranging in wavelength from long-wavelength oscillators using partial wave guiding through ultraviolet through hard x-ray that are either seeded or start from noise (SASE). In addition, FELs that produce different polarizations of the output radiation ranging from linear through elliptic to circular polarization are currently under study. In this paper, we develop a three-dimensional, time-dependent formulation that is capable of modeling this large variety of FEL configurations including different polarizations. We employ a modal expansion for the optical field, *i.e.*, a Gaussian expansion with variable polarization for free-space propagation. This formulation uses the full Newton-Lorentz force equations to track the particles through the optical and magnetostatic fields. As a result, arbitrary three-dimensional representations for different undulator configurations are implemented, including planar, helical, and elliptical undulators. In particular, we present an analytic model of an APPLE-II undulator to treat arbitrary elliptical polarizations, which is used to treat general elliptical polarizations. To model oscillator configurations, and allow propagation of the optical field outside the undulator and interact with optical elements, we link the FEL simulation with the optical propagation code OPC. We present simulations using the APPLE-II undulator model to produce elliptically polarized output radiation, and present a detailed comparison with recent experiments using a tapered undulator configuration at the Linac Coherent Light Source. Validation of the nonlinear formation is also shown by comparison with experimental results obtained in the SPARC SASE FEL experiment at ENEA Frascati, a seeded tapered amplifier experiment at Brookhaven National Laboratory, and the 10-kW Upgrade Oscillator experiment at the Thomas Jefferson National Accelerator Facility.

## I. INTRODUCTION

While free-electron lasers (FELs) have been intensively studied since the 1970s, new developments and concepts keep the field fresh. Intensive work is ongoing into new FEL-based light sources that probe ever shorter wavelengths with a variety of configurations. There presently exists a large variety of FELs ranging from long-wavelength oscillators using partial wave guiding to ultraviolet and hard x-ray FELs that are either seeded or starting from noise (*i.e.*, Self-Amplified Spontaneous Emission or SASE). As these new light sources come on-line, interest will grow in shorter pulses, new spectral ranges and higher photon fluxes. In addition, interest is growing in producing photons with a variety of polarizations ranging from linear, through elliptical, to circular. Indeed, novel configurations have been described for producing variable polarizations in synchrotron light sources and FELs using a variety of different undulator designs including APPLE-II and Delta-type undulators [1-7]. In this paper, we develop a three-dimensional, time-dependent nonlinear formulation that is capable of modeling such a large variety of FELs, in particular this represents the first presentation of a three-dimensional simulation of elliptically polarized radiation from a FEL.

We present an analytic model of an APPLE-II undulator in order to simulate elliptical polarizations. We employ a Gaussian modal expansion for the optical field. Particle dynamics are treated using the full Newton-Lorentz force equations to track the particles through the optical and magnetic fields. To allow propagation of the optical field outside the undulator and interact with optical elements, we interface with the optical propagation code OPC [8,9].

An important motivation in this development is the ability to describe the interaction in arbitrarily polarized undulators including linear, elliptical, and helical polarizations. To that end, self-consistent, three-dimensional representations of these undulator types are included in the formulation. This includes an approximate analytical model of an APPLE-II undulator.

The organization of the paper is as follows. General properties of the formulation are described in detail in Section II. The field representations used for the undulator fields, quadrupole and dipole fields, and the Gaussian optical fields are described in Section III, and the dynamical equations are discussed in Section IV. We demonstrate that the dynamical equations describe vacuum diffraction in the limit in which the electron beam vanishes in Sec. V. A simulation showing the application of the formulation for an elliptically polarized undulator is described in Section VI, and a discussion of the comparison of the simulation results with a generalization of the parameterization due to Ming Xie [10] to include the elliptical undulator is also presented. Simulations of the Linac Coherent Light Source (LCLS) at the Stanford Linear Accelerator Center (SLAC) [11,12] are presented in Sec. VII. In particular, we discuss the comparison of the simulation with the first lasing experiment [11], and then go on to compare the simulation with recent experiments on the LCLS using a strongly tapered undulator [12]. Comparisons with a variety of FEL experiments are presented in Sections VIII – XI including another SASE FEL, a seeded and tapered amplifier, and an oscillator in order to provide a more comprehensive validation of the formulation. The SPARC SASE FEL [13] conducted at ENEA Frascati is discussed in Sec. VIII. This is followed by a comparison of the simulation with a seeded, infrared, tapered-amplifier experiment [14] at Brookhaven National Laboratory in Sec. IX. Simulation of the IR-Upgrade FEL oscillator experiment [15] at the Thomas Jefferson National Accelerator Facility (JLab) is presented in Sec. X. This covers the three major configurations used in FEL experiments: SASE, oscillators, and seeded amplifiers. A summary and discussion is given in Section XI.

## II. GENERAL SIMULATION PROPERTIES

The formulation we develop describes the particles and fields in three spatial dimensions and includes time dependence as well. Electron trajectories are integrated using the complete Newton-Lorentz force equations. No wiggler-averaged-orbit approximation is made. The magnetostatic fields can be specified by analytical functions for a variety of analytic undulator models (such as planar, elliptical, or helical representations), quadrupoles, and dipoles. These magnetic field elements can be placed in arbitrary sequences to specify a variety of different transport lines. As such, we can set up field configurations for single or multiple wiggler segments with quadrupoles either placed between the undulators or superimposed upon the undulators to create a FODO lattice. Dipole chicanes can also be placed between the undulators to model various optical klystron and/or high-gain harmonic generation (HGHG) configurations. The fields can also be imported from a field map.

The electromagnetic field is described by a modal expansion. For free-space propagation, we use Gaussian optical modes. The Gauss-Hermite modes are used for simulation of planar undulators, while Gauss-Laguerre modes are used for elliptical or helical undulators.

The electromagnetic field representations are also used in integrating the electron trajectories, so that harmonic motions and interactions are included in a self-consistent way. Further, the same integration engine is used within the undulator(s) as in the gaps, quadrupoles, and dipoles, so that the phase of the optical field relative to the electrons is determined self-consistently when propagating the particles and fields in the gaps between the undulators.

Particle loading is done in a deterministic way using Gaussian quadrature that preserves a quiet start for both the fundamental and all harmonics. Shot noise is included using a Poisson statistics algorithm [16] so that the formulation is capable of simulating SASE FELs; however, provision is made for enhanced shot-noise due to various levels of micro-bunching.

The FEL simulation has also been linked to the Optics Propagation Code (OPC) [8,9] for the simulation of FEL oscillators or propagating an optical field beyond the end of the undulator line to a point of interest. OPC propagates the optical field using either the Fresnel diffraction integral or the spectral method in the paraxial approximation using fast discrete Fourier transforms (FFT). A modified Fresnel diffraction integral [17, 18] is also available and allows the use of FFTs in combination with an expanding grid on which the optical field is defined. This method is often used when diffraction of the optical field is large. Propagation can be done either in the time or frequency domain. The latter allows for the inclusion of dispersion and wavelength dependent properties of optical components. Currently, OPC includes mirrors, lenses, phase and amplitude masks, and round and rectangular diaphragms. Several optical elements can be combined to form more complex optical component, *e.g.*, by combining a mirror with a hole element, extraction of radiation from a resonator through a hole in one of the mirrors can be modelled. Phase masks can be used, for example, to model mirror distortions or to create non-standard optical components like a cylindrical lens.

In a typical resonator configuration, OPC handles the propagation from the end of the gain medium to the first optical element, applies the action of the optical element to the optical field and propagates it to the next optical element and so on until it reaches the entrance of the gain medium. Diagnostics can be performed at the planes where the optical field is evaluated. Some optical elements, specifically diaphragms and mirrors allow forking of the optical path. For example, the reflected beam of a partial transmitting output mirror forms the main intracavity optical path, while the transmitted beam is extracted from the resonator. When the intracavity propagation reaches the output mirror, this optical propagation can be temporarily suspended, and the extracted beam can be propagated to a diagnostic point for evaluation. Then the intra-cavity propagation (main path) is resumed.

The numerical procedure involves translating between the input/output required for the FEL simulation and OPC. Initially, we run the FEL simulation to determine the optical output after the first pass through the undulator, which then writes a file describing the complex field of the optical mode. OPC is then used to propagate this field to the downstream mirror, which is partially transmissive in the current example. The portion of the optical mode that is reflected is then propagated to the upstream mirror (which is a high reflector) by OPC, and then back to the undulator entrance. The field at the undulator entrance is then reduced to an ensemble of Gaussian modes that is used as input to the FEL simulation for the next pass. This process is repeated for an arbitrary number of passes. While the example discussed in this paper relates to a concentric resonator, OPC has also been used to simulate a regenerative amplifier with a ring resonator [19].

## III. THE FIELD REPRESENTATIONS

The undulator field models are three-dimensional representations. Two planar undulator models are available corresponding to flat-pole-faces and parabolic-pole-faces.

The parabolic-pole-face model provides weak two-plane focusing. The elliptical undulator field is modeled by a representation of an APPLE-II undulator consisting of two flat-pole-face undulators that are shifted in phase. In each case, however, the injection into and ejection from the undulators is simulated by the particle tracking algorithms using smooth models for the undulator transitions. The quadrupole and dipole field models used are curl- and divergence-free representations with hard-edged field transitions.

### A. The Flat-Pole-Face Undulator

The flat-pole-face undulator is represented by

$$\mathbf{B}_w(\mathbf{x}) = B_w(z)\left(\sin k_w z - \frac{\cos k_w z}{k_w B_w}\frac{dB_w}{dz}\right)\hat{e}_y \cosh k_w y$$
$$+ B_w(z)\hat{e}_z \sinh k_w y \cos k_w z, \quad (1)$$

where $B_w$ and $k_w$ (=$2\pi/\lambda_w$, where $\lambda_w$ is the undulator period) are the undulator amplitude and wavenumber respectively.

This field is both curl- and divergence-free when the amplitude, $B_w$, is constant. The transitions at the ends of each undulator segment are modeled via

$$B_w(z) = \begin{cases} B_{w0}\sin^2\left(\frac{k_w z}{4N_{tr}}\right) & ; 0 \leq z \leq N_{tr}\lambda_w \\ B_{w0}\cos^2\left(\frac{k_w(L_{tr}-z)}{4N_{tr}}\right) & ; L_{tr} \leq z \leq L_w \end{cases}, \quad (2)$$

where $B_{w0}$ is the field amplitude in the uniform region, $L_w$ is the undulator segment length, $N_{tr}$ is the number of undulator periods in the transition region, and $L_{tr}$ (= $L_w - N_{tr}\lambda_w$) is the start of the output transition. The field in the transitions is divergence-free, and the $z$-component of the curl also vanishes. The transverse components of the curl do not vanish, but are of the order of $(k_w B_w)^{-1} dB_w/dz$, which are usually small.

### B. The Parabolic-Pole-Face Undulator

The parabolic-pole-face field model is given by

$$\mathbf{B}_w(\mathbf{x}) = B_w(z)\left(\cos k_w z + \frac{\sin k_w z}{k_w B_w}\frac{dB_w}{dz}\right)\hat{e}_\perp(x,y)$$
$$- \sqrt{2}B_w \hat{e}_z \cosh\left(\frac{k_w x}{\sqrt{2}}\right)\sinh\left(\frac{k_w y}{\sqrt{2}}\right)\sin k_w z, \quad (3)$$

where

$$\hat{e}_\perp(x,y) = \hat{e}_x \sinh\left(\frac{k_w x}{\sqrt{2}}\right)\sinh\left(\frac{k_w y}{\sqrt{2}}\right)$$
$$+ \hat{e}_y \cosh\left(\frac{k_w x}{\sqrt{2}}\right)\cosh\left(\frac{k_w y}{\sqrt{2}}\right), \quad (4)$$

and $B_w(z)$ is given in Eq. (2). As in the case of the flat-pole-face model, this field is divergence-free and the $z$-component of the curl also vanishes.

### C. The Helical Undulator

The helical undulator model that is employed is of the form in cylindrical coordinates

$$\mathbf{B}_w(\mathbf{x}) = 2B_w(z)\left(\cos \chi - \frac{\sin \chi}{k_w B_w}\frac{dB_w}{dz}\right)I_1'(k_w r)\hat{e}_r$$
$$- 2B_w(z)\left(\sin \chi + \frac{\cos \chi}{k_w B_w}\frac{dB_w}{dz}\right)\frac{1}{k_w r}I_1(k_w r)\hat{e}_\theta$$
$$+ 2B_w(z)I_1(k_w r)\hat{e}_z \sin \chi, \quad (5)$$

where $\chi = k_w z - \theta$, $I_1$ denotes the regular Bessel function of the first kind, and $B_w(z)$ is given by Eq. (2).

### D. The APPLE-II Undulator Model

An approximate representation of an APPLE-II undulator can be formed by the super-position of two flat-pole-face undulator models that are oriented perpendicularly to each other and phase shifted with respect to the axis of symmetry. As such, the field is represented in the form

$$\mathbf{B}_w(\mathbf{x}) = B_w(z)\left(\sin(k_w z + \phi) - \frac{\cos(k_w z + \phi)}{k_w B_w}\frac{dB_w}{dz}\right)$$
$$\times \hat{e}_x \cosh k_w x$$
$$+ B_w(z)\left(\sin k_w z - \frac{\cos k_w z}{k_w B_w}\frac{dB_w}{dz}\right)\hat{e}_y \cosh k_w y$$
$$+ B_w(z)\hat{e}_z \sinh k_w x[\cos(k_w z + \phi) + \cos k_w z], \quad (6)$$

where, as before, $B_w(z)$ is given by Eq. (2). This is an approximate representation of an APPLE-II undulator that is valid near the axis of symmetry. The ellipticity is governed by the choice of the phase, $\phi$.

For $0 \leq \phi \leq \pi/2$, the ellipticity, $u_e$, is given by

$$u_e = \frac{1 - \cos \phi}{1 + \cos \phi}, \quad (7)$$

for which the semi-major axis is oriented along $\pi/2$. The choice of $\phi = 0$ ($\pi/2$) corresponds to planar (helical) polarization. When $\pi/2 \leq \phi \leq \pi$, the ellipticity is

$$u_e = \frac{1 + \cos \phi}{1 - \cos \phi}, \quad (8)$$

and the semi-major axis is oriented along $-\pi/2$.

Illustrations of the on-axis field contours are shown in Fig. 1, where we plot the $y$-component of the field versus the $x$-component (normalized to the amplitude) for $\phi = \pi/8$, $\pi/4$, $\pi/2$, and $3\pi/4$.

The choice of elliptical polarization for the Gaussian modes has the semi-major axis aligned along the *x*-axis, so that this undulator field must be rotated in order to correspond to the polarization of the radiation field.

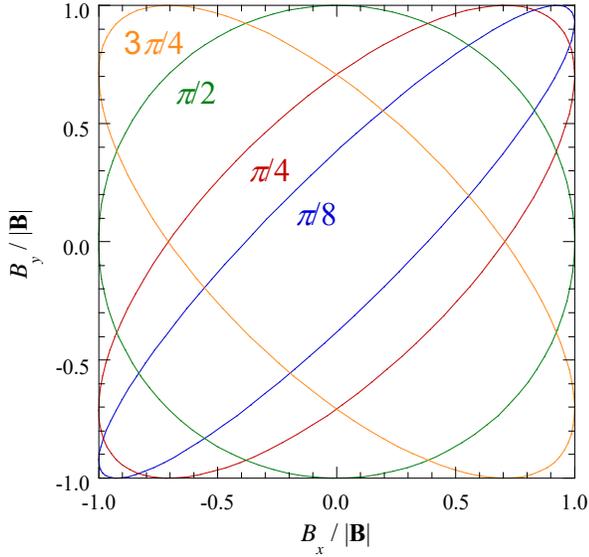

Fig. 1: The on-axis field contours for different phases $\phi = \pi/8$, $\pi/4$, $\pi/2$, and $3\pi/4$.

### E. Quadrupole and Dipole Fields

The quadrupole field model used is

$$\mathbf{B}_Q(\mathbf{x}) = B_Q(z)(y\hat{\mathbf{e}}_x + x\hat{\mathbf{e}}_y) \ , \quad (9)$$

where $B_Q(z)$ is the field gradient (constant) defined over a range $z_1 \leq z \leq z_2$. This field is both curl- and divergence-free over this range.

The dipole field model is described by a constant field oriented perpendicularly to the axis of symmetry over some range $z_1 \leq z \leq z_2$.

### F. The Gaussian Optical Modes

The Gauss-Hermite modes are used in simulating the interaction with planar undulators. In this case, the field representation is

$$\delta\mathbf{A}(\mathbf{x},t) = \hat{\mathbf{e}}_x \sum_{\substack{l,n,=0 \\ h=1}}^{\infty} e_{l,n,h}\left(\delta A_{l,n,h}^{(1)} \sin \varphi_h + \delta A_{l,n,h}^{(2)} \cos \varphi_h\right), \quad (10)$$

where the indices $(l,n)$ is describe the transverse mode structure, the index $h$ is the harmonic number, the field amplitudes, $\delta A_{l,n,h}^{(1,2)}$, vary slowly in $(z,t)$,

$$e_{l,n,h} = \frac{w_{0,h}}{w_h} \exp\left(-r^2/w_h^2\right) H_l\left(\frac{\sqrt{2}x}{w_h}\right) H_n\left(\frac{\sqrt{2}y}{w_h}\right), \quad (11)$$

describes the transverse mode structure where $H_{l,n}$ are the Hermite polynomials, $w_{0,h}$ and $w_h$ denote the waist size and spot size of the $h^\text{th}$ harmonic respectively. The spot size is assumed to be a slowly-varying function of $(z,t)$. The phase is

$$\varphi_h = h(k_0 z - \omega t) + \alpha_h \frac{r^2}{w_h^2} \ , \quad (12)$$

where $k_0 = \omega/c$, $\alpha_h$ denotes the curvature of the phase front of the $h^\text{th}$ harmonic and which is assumed to be a slowly-varying function of $(z,t)$.

The Gauss-Laguerre modes are used when simulating elliptical and helical undulators. The field representation is

$$\delta\mathbf{A}(\mathbf{x},t) = \sum_{\substack{l=-\infty \\ n=0, h=1}}^{\infty} e_{l,n,h}\{\delta A_{l,n,h}^{(1)}(\hat{\mathbf{e}}_x \sin \varphi_{l,h} \pm u_e\hat{\mathbf{e}}_y \cos \varphi_{l,h}) \\ + \delta A_{l,n,h}^{(2)}(\hat{\mathbf{e}}_x \cos \varphi_{l,h} \mp u_e\hat{\mathbf{e}}_y \sin \varphi_{l,h})\}, \quad (13)$$

where the transverse mode structure is given by

$$e_{l,n,h} = \frac{w_{0,h}}{w_h} \exp\left(-r^2/w_h^2\right) \zeta^{|l|} L_n^{|l|}(\zeta^2), \quad (14)$$

$L_n^l$ is the associated Laguerre polynomial, and $\zeta = \sqrt{2}r/w_h$. The phase is given by

$$\varphi_{l,h} = h(k_0 z - \omega t) + l\theta + \alpha_h \frac{r^2}{w_h^2} \ . \quad (15)$$

The total power carried in each mode, $P_{l,n,h}$, is given by integration of the Poynting vectors over the cross section. This is given by

$$P_{l,n,h} = \frac{m_e^2 c^5}{8e^2} 2^{l+n-1} l! n! k_0^2 w_{0,h}^2 \left(\delta a_{l,n,h}^{(1)\,2} + \delta a_{l,n,h}^{(2)\,2}\right), \quad (16)$$

for the Gauss-Hermite modes, and

$$P_{l,n,h} = \frac{m_e^2 c^5}{8e^2} \frac{(|l|+n)!}{n!} k_0^2 w_{0,h}^2 \left(\delta a_{l,n,h}^{(1)\,2} + \delta a_{l,n,h}^{(2)\,2}\right), \quad (17)$$

for the Gauss-Laguerre modes, where $\delta a_{l,n,h}^{(1,2)}$ ($= e\delta A_{l,n,h}^{(1,2)}/m_e c^2$) is the normalized field amplitude, and $m_e^2 c^5/8e^2 \cong 1.089$ GW.

## IV. THE DYNAMICAL EQUATONS

The dynamical equations for the fields employ the Source-Dependent Expansion [20] which is an adaptive eigenmode algorithm in which the evolution of the spot size and curvature are determined self-consistently in terms of the interaction with the electron beam. As such, the dynamical equations for the fields are of the form

$$\frac{d}{dz}\begin{pmatrix} \delta a_{l,n,h}^{(1)} \\ \delta a_{l,n,h}^{(2)} \end{pmatrix} + K_{l,n,h} \begin{pmatrix} \delta a_{l,n,h}^{(2)} \\ -\delta a_{l,n,h}^{(1)} \end{pmatrix} = \begin{pmatrix} S_{l,n,h}^{(1)} \\ S_{l,n,h}^{(2)} \end{pmatrix}, \quad (18)$$

where $S_{l,n,h}^{(1,2)}$ are the source terms,

$$\frac{d}{dz} = \frac{\partial}{\partial z} + \frac{1}{c}\frac{\partial}{\partial t} \ , \quad (19)$$

is the convective derivative, and

$$K_{l,n,h} = F_{l,n} \frac{w_{0,h}^2}{w_h^2}\left(\frac{\alpha_h}{w_h}\frac{dw_h}{dz} - \frac{1}{2}\frac{d\alpha_h}{dz} - \frac{1+\alpha_h^2}{k_0 w_h^2}\right), \quad (20)$$

for $F_{l,n} = 1 + l + n$ ($= 1 + |l| + 2n$) for the Gauss-Hermite (Gauss-Laguerre) modes. The source terms are given by

$$\begin{pmatrix} S^{(1)}_{l,n,h} \\ S^{(2)}_{l,n,h} \end{pmatrix} = \frac{\omega_b^2}{k_0 c^2} \frac{1}{2^{l+n-1} \pi\, l!\, n!} \frac{1}{w_{0,h}^2}$$
$$\times \left\langle e_{l,n,h} \frac{v_x}{|v_z|} \begin{pmatrix} -\cos\varphi_h \\ \sin\varphi_h \end{pmatrix} \right\rangle , \quad (21)$$

for the Gauss-Hermite modes, and

$$\begin{pmatrix} S^{(1)}_{l,n,h} \\ S^{(2)}_{l,n,h} \end{pmatrix} = \frac{\omega_b^2 c^5}{k_0 c^2} \frac{n!}{(|l|+n)!} \frac{1}{w_{0,h}^2}$$
$$\times \left\langle \frac{e_{l,n,h}}{|v_z|} \begin{pmatrix} v_x \cos\varphi_{l,h} \mp u_e v_y \sin\varphi_{l,h} \\ -v_x \sin\varphi_{l,h} \mp u_e v_y \cos\varphi_{l,h} \end{pmatrix} \right\rangle , \quad (22)$$

for the Gauss-Laguerre modes, where $\omega_b$ is the beam plasma frequency and $\langle(...)\rangle$ denotes an average over the initial beam distribution. A uniform distribution in initial phase and a Gaussian distribution in coordinate and momentum space is assumed in the examples discussed in this paper. In this case

$$\langle(\cdots)\rangle = \int_0^{2\pi} \frac{d\psi_0}{2\pi} \int_1^\infty \frac{d\gamma_0}{\sqrt{\pi/2}\,\Delta\gamma} \exp\left[-(\gamma_0 - \gamma_{avg})^2/2\Delta\gamma^2\right]$$
$$\times \iint \frac{dx_0 dy_0}{2\pi\sigma_r^2} \iint \frac{dp_{x0} dp_{y0}}{2\pi\sigma_p^2} \exp\left(-r_0^2/2\sigma_r^2 - p_{\perp 0}^2/2\sigma_p^2\right)(\cdots)$$
(23)

where $\gamma_{avg}$ and $\Delta\gamma$ denote the average energy and energy spread, and $\sigma_r$ and $\sigma_p$ describe the initial transverse phase space.

The evolution of the spot size and curvature are governed by

$$\frac{dw_h}{dz} = \frac{2\alpha_h}{k_0 w_h} - w_h Y_h , \quad (24)$$

$$\frac{1}{2}\frac{d\alpha_h}{dz} = \frac{1 + \alpha_h^2}{k_0 w_h^2} - X_h - \alpha_h Y_h , \quad (25)$$

where the source terms are defined as

$$X_h = -2\frac{\left(S^{(2)}_{0,2,h} + S^{(2)}_{2,0,h}\right)\delta a^{(1)}_{0,0,h} - \left(S^{(1)}_{0,2,h} + S^{(1)}_{2,0,h}\right)\delta a^{(2)}_{0,0,h}}{\delta a^2_{0,0,h}} , \quad (26)$$

$$Y_h = -2\frac{\left(S^{(1)}_{0,2,h} + S^{(1)}_{2,0,h}\right)\delta a^{(1)}_{0,0,h} + \left(S^{(2)}_{0,2,h} + S^{(2)}_{2,0,h}\right)\delta a^{(2)}_{0,0,h}}{\delta a^2_{0,0,h}} , \quad (27)$$

for the Gauss-Hermite modes, and

$$X_h = -\frac{S^{(1)}_{0,1,h}\delta a^{(2)}_{0,0,h} - S^{(2)}_{0,1,h}\delta a^{(1)}_{0,0,h}}{\delta a^2_{0,0,h}} , \quad (28)$$

$$Y_h = \frac{S^{(1)}_{0,1,h}\delta a^{(1)}_{0,0,h} + S^{(2)}_{0,1,h}\delta a^{(2)}_{0,0,h}}{\delta a^2_{0,0,h}} , \quad (29)$$

for the Gauss-Laguerre modes, where $\delta a^2_{0,0,h} = \delta a^{(1)\,2}_{0,0,h} + \delta a^{(2)\,2}_{0,0,h}$.

These field equations are integrated together with the Newton-Lorentz force equations for the particles.

$$v_z \frac{dx}{dz} = v_x , \quad (30)$$

$$v_z \frac{dy}{dz} = v_y , \quad (31)$$

$$\frac{d\psi}{dz} = k + k_w - \frac{\omega}{v_z} , \quad (32)$$

where $\psi$ is the ponderomotive phase,

$$v_z \frac{d}{dz}\mathbf{p} = -e\delta\mathbf{E} - \frac{e}{c}\mathbf{v}\times\left(\mathbf{B}_{static} + \delta\mathbf{B}\right) , \quad (33)$$

where $\delta\mathbf{E}$ and $\delta\mathbf{B}$ correspond to the electric and magnetic fields of the complete super-position of Gaussian modes, and $\mathbf{B}_{static}$ is the magnetostatic fields (undulators, quadrupoles, and dipoles).

The time dependence is treated by allowing the field *slices* to advance relative to the electron *slices* at arbitrary integration intervals. Since the optical field slips ahead of the electrons at the rate of one wavelength per undulator period, if this slippage operation is performed at shorter intervals, then the field advance is interpolated between adjacent temporal *slices* based on this slippage rate.

The total number of equations in each simulation is

$$N_{equations} = N_{slices}\left[6N_{particles} + 2\left(N_{modes} + N_{harmonics}\right)\right], \quad (34)$$

where $N_{slices}$ is the number of *slices* in the simulation, and for each *slice*, $N_{particles}$ is the number of particles, $N_{modes}$ is the number of modes in all the harmonics, and $N_{harmonics}$ is the number of harmonics. This complete set of coupled nonlinear differential equations is solved numerically using a Runge-Kutta algorithm. The particle averages in the source terms are implemented by converting the continuous integral over a distribution function into a discrete set of macro-particles using Gaussian quadrature over each of the degrees of freedom. Since the Newton-Lorentz equations are integrated for each macro-particle, the step size must be small enough to resolve the wiggle-motion in the undulators. In practical terms, this means that simulations must take 20 or more steps per undulator period. However, the Runge-Kutta algorithm allows for changing the step size "on the fly", and longer integration steps are used in the drift spaces between undulator segments.

## V. VACUUM PROPAGATION

That these equations recover vacuum propagation can be demonstrated by considering the case in which the electron beam is not present and the sources [Eqs. (21) and (22)] vanish. As a result, $X_h = Y_h = 0$ so that the spot size and curvature satisfy the following equations

$$\frac{dw_h}{dz} = \frac{2\alpha_h}{k_0 w_h} , \quad (35)$$

and

$$\frac{1}{2}\frac{d\alpha_h}{dz} = \frac{1 + \alpha_h^2}{k_0 w_h^2} . \quad (36)$$

These equations have the well-known solutions for the spot size and curvature *in vacuo* where

$$w_h(z) = w_{0,h}\sqrt{1 + \frac{(z - z_0)^2}{z_R^2}} , \quad (37)$$

and

$$\alpha_h(z) = \frac{z - z_0}{z_R} , \quad (38)$$

where $z_R = k_0 w_{0,h}^2/2$ is the Rayleigh range, and $z_0$ denotes the position of the mode waist.

Substitution of Eqs. (35) and (36) into Eq. (20) shows that

$$K_{l,n,h} = -\frac{w_{0,h}^2}{w_h^2}\frac{F_{l,n}}{z_R} , \quad (39)$$

which is the derivative of the Gouy phase shift, $\varphi_{l,n,h}$. If we now express the field components in the form $\delta A_{l,n,h}^{(1)} = \delta A_{l,n,h}\cos\varphi$ and $\delta A_{l,n,h}^{(2)} = \delta A_{l,n,h}\sin\varphi$, then the dynamical equations can be reduced to equations for the derivatives of the amplitude $\delta A_{l,n,h}$ and phase $\varphi$ as

$$\frac{d}{dz}\delta A_{l,n,h} = 0 , \quad (40)$$

which shows that the power is constant, and

$$\frac{d}{dz}\varphi = \frac{d}{dz}\varphi_{l,n,h} , \quad (41)$$

which indicates that the phase variation is described by the Gouy phase.

As a result, the dynamical equations describe vacuum diffraction in the absence of an electron beam.

## VI. SIMULATION OF ELLIPTICAL UNDULATORS

We now describe the generation of elliptically polarized radiation using an elliptically polarized undulator. For convenience, we consider the same beam, undulator and focusing configuration as used in the simulation of the SPARC experiment (Sec. VIII), except that we now use the APPLE-II undulator model and elliptically polarized radiation. In addition, we limit the simulation to the steady-state (*i.e.*, a single temporal slice) regime since that is sufficient to demonstrate the reliability of the formulation and allows us to compare the simulation results with an analytic theory.

In order to compare the simulation results with an analytic theory, we make use of a description of the effect of an elliptical undulator on the resonant wavelength and the usual *JJ*-coupling factor that has been given by J.R. Henderson *et al.* [21]. The generalized resonance condition varies with the ellipticity as follows

$$\lambda = \frac{\lambda_w}{2\gamma^2}\left[1 + (1 + u_e^2)\frac{K^2}{2}\right]. \quad (42)$$

Observe that this reduces to the usual expressions in the limits of planar ($u_e = 0$) and helical ($u_e = 1$) undulators. The generalized *JJ*-factor is given by

$$JJ = \sqrt{1 + u_e^2}\,\frac{K}{\sqrt{2}}\left[J_0(\zeta) - \frac{(1 - u_e^2)}{(1 + u_e^2)}J_1(\zeta)\right], \quad (43)$$

where

$$\zeta = \frac{(1 - u_e^2)K^2/4}{1 + (1 + u_e^2)K^2/2} . \quad (44)$$

In ref. [21], the authors compared the results of simulations for different choices of the ellipticity using (1) a one-dimensional, orbit-averaged simulation code in which the generalized resonance condition and *JJ*-factor were implemented, and (2) the implementation of an elliptical undulator model in the one-dimensional particle-in-cell PUFFIN [22] code. Since the PUFFIN code does not make use of the orbit average and does not explicitly include either the resonance condition or the *JJ*-factor, it is expected that the ellipticity is included self-consistently. The comparison of the two codes showed excellent agreement. Hence, we conclude that the generalized dynamical equations constitute a reliable description of the ellipticity. As a result, we can obtain a three-dimensional approximation of the interaction in an elliptical undulator by using these expressions for the resonant wavelength and *JJ*-factor in the parameterization given by Ming Xie [10]. This generalized parameterization is then compared with the results of three-dimensional simulations.

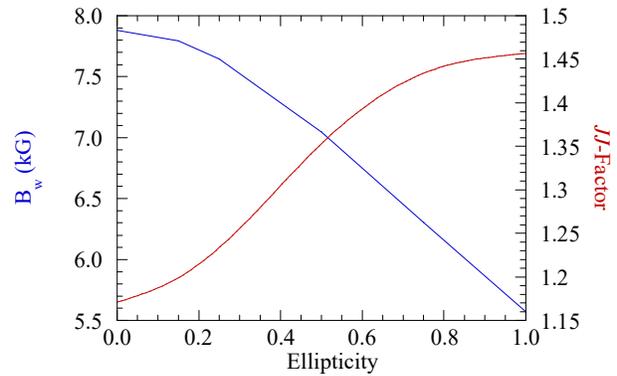

Fig. 2: The generalized resonant undulator field (left) and *JJ*-factor (right) versus the ellipticity.

The undulator field amplitude (left axis in blue) associated with the generalized resonance and the generalized *JJ*-factor (right axis in red) for the parameters of interest are shown in Fig. 2 as functions of the ellipticity. We employed these undulator field amplitudes in performing the simulations for various choices of the ellipticity. Note, however, that the simulation model does not employ a wiggler-averaged orbit integration; hence, the physics associated with the *JJ*-factor is implicitly included in the simulations. As a result, the *JJ*-factor is only used in generalizing the parameterization developed by Ming Xie for comparison purposes.

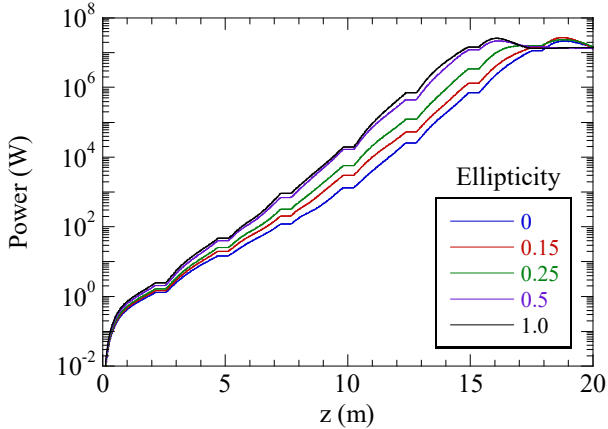

Fig. 3: The power along the undulator for various choices of the ellipticity.

Simulations have been performed for ellipticities ranging from zero (planar undulator) to unity (helical undulator) using the APPLE-II undulator representation. In each case, the simulation was started from shot noise using the same noise seed. No average over multiple noise seeds was performed; however, this is perfectly adequate since our intention is to study the variation in performance due to different ellipticities and the initial phase space used in the different simulations is invariant with respect to the ellipticity. Results showing the power growth along the undulator line are shown in Fig. 3 for ellipticities of 0, 0.15, 0.25, 0.50 and 1.0. As shown in the figure, the distance to saturation tends to decrease with increasing ellipticity. This is understandable since the *JJ*-factor increases with the ellipticity and this tends to increase the strength of the interaction.

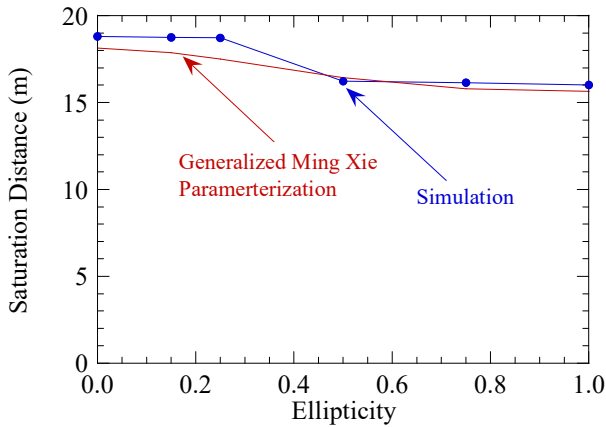

Fig. 4: Variation in the distance to saturation versus the ellipticity.

A comparison between the saturation distances found in simulations and the predictions based on the generalized parameterization due to Ming Xie is shown in Fig. 4 where we plot the saturation distance versus the ellipticity. It should be remarked that we have added the drift space between the undulators to the predictions of the generalized parameterization. Since the simulation includes two extra undulator periods in each undulator to model the transitions at the entrances and exits of the undulators, we have added these lengths to the generalized parameterization as well. It is evident from the figure that the simulation is in good agreement with the generalized parameterization.

### VII. THE LCLS SASE FEL

The LCLS [11] is a SASE FEL user facility that became operational in 2009 operating at a 1.5 Å wavelength. In this paper, we first discuss a comparison with the first lasing results from the LCLS in order to validate the model. We then present the first comparison showing substantial agreement between simulation and an experiment on the LCLS that employed an aggressive taper to enhance the efficiency.

| **Electron Beam** | |
|---|---|
| Energy | 13.64 GeV |
| Bunch Charge | 250 pC |
| Bunch Duration | 83 fsec |
| Peak Current | 3000 A (flat-top) |
| *x*-Emittance | 0.4 mm-mrad |
| *y*-Emittance | 0.4 mm-mrad |
| rms Energy Spread | 0.01% |
| rms Size ($x$) | 21.5 microns |
| $\alpha_x$ | 1.1 |
| $\beta_x$ | 30.85 m |
| rms Size ($y$) | 19.5 microns |
| $\alpha_y$ | -0.82 |
| $\beta_y$ | 25.38 m |
| **Undulators** | 33 segments |
| Period | 3.0 cm |
| Length | 113 Periods |
| Amplitude (1st segment) | 12.4947 kG |
| $K_{rms}$ (1st segment) | 2.4748 |
| Taper Slope | -0.0016 kG |
| Gap Length | 0.48 m |
| **Quadrupoles** | |
| Length | 7.4 cm |
| Field Gradient | 4.054 kG/cm |

Table 1: Parameters of the LCLS FEL experiment

The fundamental operating parameters are listed in Table 1. It employs a 13.64 GeV/250 pC electron beam with a flat-top temporal pulse shape of 83 fsec duration. The normalized emittance (*x* and *y*) is 0.4 mm-mrad and the rms energy spread is 0.01%. The undulator line consisted of 33 segments with a period of 3.0 cm and a length of 113 periods including one period each in entry and exit tapers. A mild down-taper in field amplitude of −0.0016 kG/segment starting with the first segment (with an amplitude of 12.4947 kG and $K_{rms}$ = 2.4748) and continuing from segment to segment was used. This is the so-called *gain taper*. The electron beam was matched into a FODO lattice consisting of 32 quadrupoles each having a field gradient of 4.054 kG/cm and a length of 7.4 cm. Each quadrupole was placed a distance of 3.96 cm downstream from the end of the preceding undulator segment. The Twiss parameters for this FODO lattice are also shown in Table 1.

The propagation of the beam through the LCLS undulator/quadrupole lattice as found in simulation is shown in Fig. 5, where we plot the beam envelope in $x$ (blue, left axis) and $y$ (red, right axis) versus position. Observe that the beam is well-confined over the 130 meters of the extended lattice with an average beam size of about 21 microns.

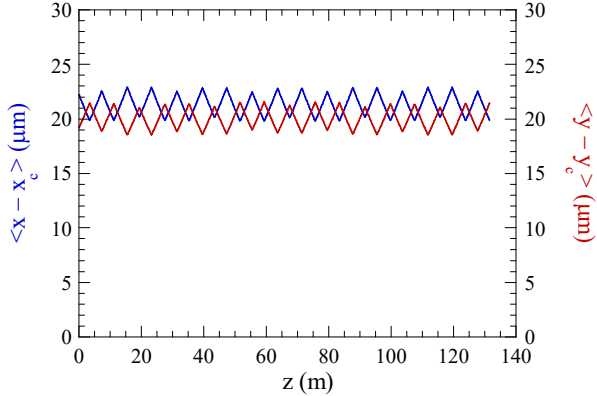

Fig. 5: Simulated propagation of the LCLS beam.

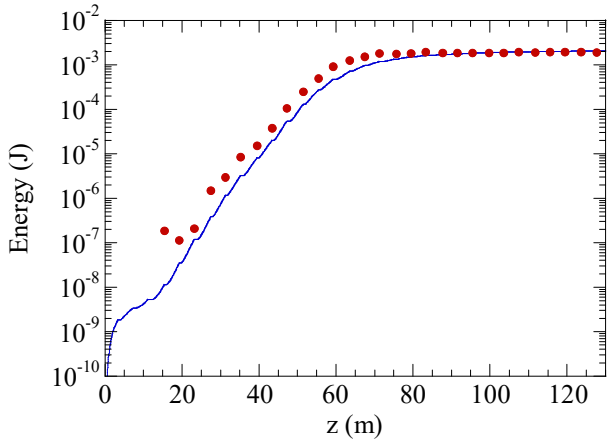

Fig. 6: Comparison between the simulation and experimental data (red circles) from the LCLS (courtesy of P. Emma and H.-D. Nuhn) and simulation (blue) using the gain taper.

The LCLS produces pulses of about 1.89 mJ at the end of the undulator line [11], and saturation is found after about 65 – 75 m along the undulator line. A comparison between the measured pulse energies (red circles) and the simulation (blue) is shown in Fig. 6. The experimental data is courtesy of P. Emma and H.-D. Nuhn at SLAC, and the simulation results represent an average over an ensemble of 25 runs performed with different noise seeds. As shown in the figure, the simulations are in good agreement with the measurements in the start-up and exponential growth regions. The simulation exhibits saturation at the same distance as the experiment in the range of 65 – 75 m at a pulse energy of 1.5 mJ. After saturation, in view of the gain taper, the pulse energy grows more slowly to about 2.02 mJ at the end of the undulator line, which is approximately 8% higher than the observed pulse energy.

Experiments have also been performed at the LCLS [12] to investigate enhancing the efficiency using a more sharply tapered undulator. The LCLS configured with a stronger taper for the last segments has demonstrated enhancements in the efficiency. This experiment employed an undulator in which the aforementioned mild linear down-taper is enhanced by the addition of a more rapid down-taper starting at the 14[th] undulator segment. This so-called saturation taper profile is shown in Fig. 7 (data courtesy of D. Ratner).

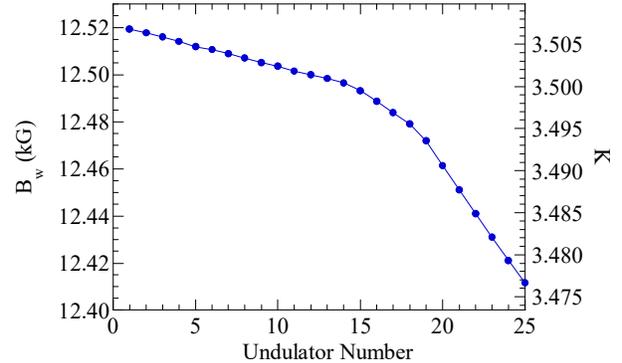

Fig. 7: The experimentally applied saturation taper profile.

In comparison with the undulator and electron beam properties employed in the first lasing experiments, the tapered undulator experiment employed undulators tuned to somewhat different field strengths and electron beam parameters that may have varied from the first lasing experiment. The pulse energies in the experiment were obtained by measuring the energy loss in the electron beam. Simulations were conducted over a parameter range including emittances of 0.40 mm-mrad – 0.45 mm-mrad and energy spreads of 0.010% – 0.015% that are thought to characterize the electron beam.

A comparison between the measured pulse energies and simulations over the parameter range that most closely agree with the experiment is shown in Fig. 8, where the experimental results are shown in red. The maximum pulse energy shown represents an enhancement of the efficiency by a factor of 2 – 3 over what is found with the gain taper alone. As is evident from the figure, the simulations for the three choices are all very similar and are in good agreement with the measurements, indicating that the efficiency enhancement could be achieved for a variety of electron beam parameters.

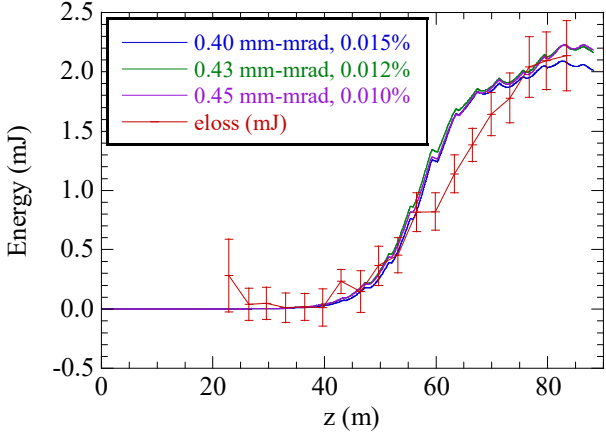

Fig. 8: Comparison between the experimental (red) and simulations for a variety of emittances and energy spreads for the saturation taper. Data courtesy of D. Ratner.

| Electron Beam | |
|---|---|
| Energy | 151.9 MeV |
| Bunch Charge | 450 pC |
| Bunch Duration | 12.67 psec |
| Peak Current | 53 A (parabolic) |
| $x$-Emittance | 2.5 mm-mrad |
| $y$-Emittance | 2.9 mm-mrad |
| rms Energy Spread | 0.02% |
| rms Size ($x$) | 132 microns |
| $\alpha_x$ | 0.938 |
| rms Size ($y$) | 75 microns |
| $\alpha_y$ | -0.705 |
| **Undulators** | 6 segments |
| Period | 2.8 cm |
| Length | 77 Periods |
| Amplitude | 7.8796 kG |
| $K_{rms}$ | 1.457 |
| Gap Length | 0.40 m |
| **Quadrupoles** | Centered in Gaps |
| Length | 5.3 cm |
| Field Gradient | 0.9 kG/cm |

Table 2: Parameters of the SPARC FEL experiment.

## VIII. THE SPARC SASE FEL

The "Sorgente Pulsata ed Amplificata di Radiazione Coerente" (SPARC) experiment is a SASE FEL located at ENEA Frascati [13]. The parameters of the experiment are summarized in Table 2 and are as follows. The electron beam energy was 151.9 MeV, with a bunch charge of 450 pC, and a bunch width of 12.67 psec. The peak current was approximately 53 A for a parabolic temporal bunch profile. The $x$ and $y$ emittances were 2.5 mm-mrad and 2.9 mm-mrad respectively, and the rms energy spread was 0.02%. There were six undulators each of which was 77 periods in length (with one period for the entrance up-taper and another for the exit down-taper) with a period of 2.8 cm and an amplitude of 7.88 kG. The gap between the undulators was 0.4 m in length and the quadrupoles (0.053 m in length with a field gradient of 0.9 kG/cm) forming a strong focusing lattice were located 0.105 m downstream from the exit of the previous undulator. Note that the quadrupole orientations were fixed and did not alternate. The electron beam was matched into the undulator/focusing lattice. The resonance occurred at a wavelength of 491.5 nm. The pulse energies were measured in the gaps between the undulator segments.

Given the bunch charge available, the SASE interaction was unable to reach saturation over the six undulators present. Hence, for the purposes of the simulation we shall add two extra undulators to bring the interaction to saturation.

The propagation of the beam through the undulator/quadrupole lattice as found in simulation is shown in Fig. 9, where we plot the beam envelope in $x$ (blue, left axis) and $y$ (red, right axis) versus position. Observe that the beam is well-confined over the 20 meters of the extended lattice with an average beam size of about 115 microns.

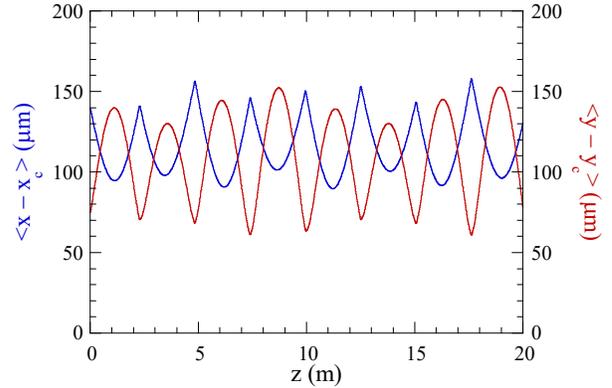

Fig. 9: Simulated propagation of the beam.

A comparison of the evolution of the pulse energy as found in simulation and as measured in the experiment is shown in Fig. 10 where the simulation is indicated by the blue line and is an average taken over 20 simulation runs with different noise seeds. The pulse energy was measured in the gaps between the undulators, and the results for a sequence of shots are indicated by the red markers (data courtesy of L. Giannessi). Observe that the agreement between the simulation and the measured performance is excellent over the entire range of the experiment. In addition, the simulation shows that saturation could have been reached after about 18 – 20 m with two additional undulator segments.

This result is in substantial agreement with the parameterization developed by Ming Xie [10]. Using a $\beta$-function of about 2 m, we find that the Pierce parameter $\rho \approx 2.88 \times 10^{-3}$ and that this parameterization predicts a gain length of 0.67 m, and a saturation distance of 18.1 m (including the additional 3.2 m represented by the gaps between the undulators). This is in reasonable agreement with the simulation.

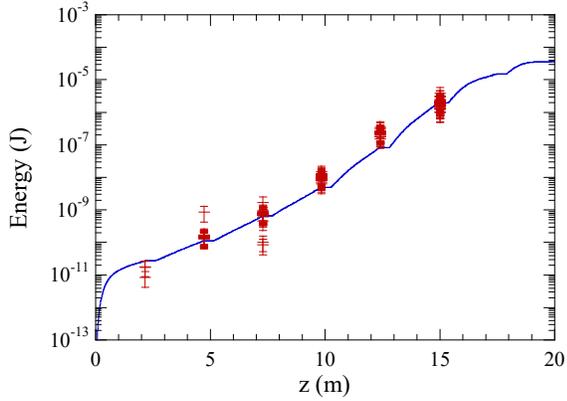

Fig. 10: Comparison of simulation results and the measured pulse energies versus distance (data courtesy of L. Giannessi).

A comparison between the evolution of the relative linewidth as determined from simulation and by measurement (data courtesy of L. Giannessi) is shown in Fig. 11 over the range of the installed undulators and agreement between the simulation and the measured linewidth is within about 35% after 15 m. As shown in the figure, the predicted linewidths are in substantial agreement.

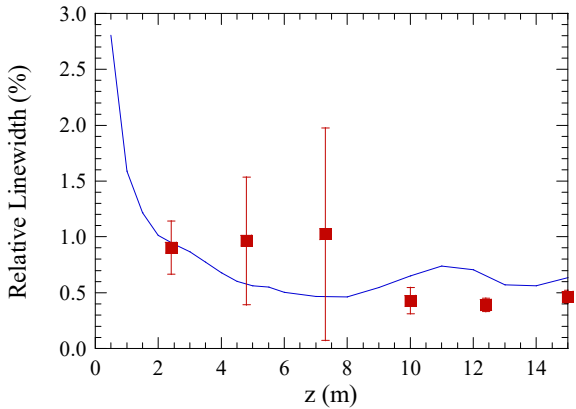

Fig. 11: Comparison of the measured relative linewidth in red (data courtesy of L. Giannessi) with that found in simulation (blue).

The initial decrease in linewidth shown in Fig. 11 results from the development of temporal coherence as can be seen in Figs. 12 – 14, where we plot the power versus time within the optical pulse. The time window used in the simulation was chosen to be 14 psec in order to allow for slippage across the 12.67 psec electron bunch. The optical pulse at the start-up of the SASE interaction is expected to contain a large number of "spikes". This is indeed what is found in simulation as shown in Fig. 15, where we plot the power in the pulse over the entire time window. This pulse is near the start of the undulator line and exhibits a broad distribution of spikes coinciding roughly with the center of the electron bunch, which is located at the center of the time window. As shown in Fig. 12, the linewidth narrows as the interaction proceeds and this corresponds to the development of temporal coherence.

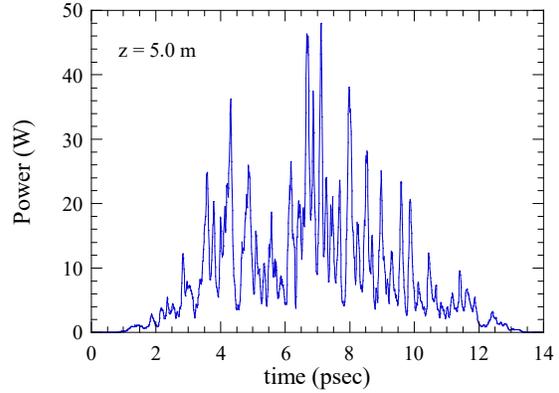

Fig. 12: Temporal pulse shape at $z = 5.0$ m.

This evolution of temporal coherence is illustrated in Figs. 13 and 14, by comparison with Fig. 12, where the temporal pulses are shown at $z = 10.0$ m and 15.0 m respectively. These two figures correspond to the exponential gain region prior to saturation. It is clear in these figures that the early collection of a large number of spikes has coalesced into a more sharply peaked distribution containing a smaller number of spikes. This corresponds to the narrowing of the linewidth due to the development of coherence in the exponential gain region.

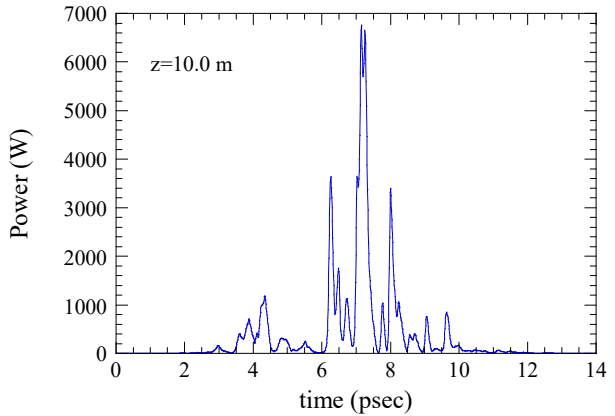

Fig. 13: Temporal pulse shape at $z = 10.0$ m.

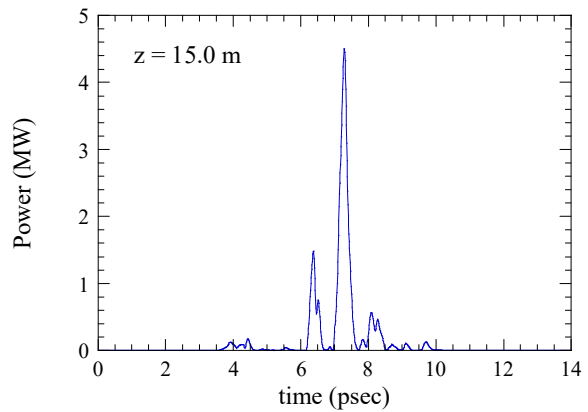

Fig. 14: Temporal pulse shape at $z = 15.0$ m.

## IX. THE BNL TAPERED AMPLIFIER

A tapered-wiggler, seeded amplifier experiment was conducted at the Source Development Laboratory at BNL [14] using a high brightness electron injector, a chicane bunch compressor feeding a 100 MeV, S-band SLAC type traveling wave linac. The electron beam is then injected into the NISUS wiggler [23] that was built for Boeing Aerospace. The NISUS wiggler is a 10 m long planar wiggler with a period of 3.89 cm and with weak, two-plane focusing. The NISUS undulator consists of a linkage of 1 meter segments, and a taper can be imposed by choosing a segment and opening the jaws of the undulator starting at that point. This creates a linear downward taper of the field. A Ti:sapphire laser was used both as the driver for the photo-cathode electron gun and as the seed laser for the FEL amplifier operating at a wavelength of 793.5 nm. A 300% enhancement over the uniform wiggler interaction was observed when the NISUS undulator was tapered.

The experimental parameters are given in Table 3. The resonant electron beam energy is 100.86 MeV, and the bunch charge is 360 pC over a bunch duration of 1.8 psec (full width), yielding a peak current of 300 A for a parabolic pulse shape. The normalized emittance was 4.0 mm-mrad and the rms energy spread was 0.1%. The electron beam was matched into the weak focusing, NISUS undulator with a matched beam radius of about 212 microns and a 2.23 m $\beta$-function. The amplitude of the NISUS wiggler in the uniform section was 3.03 kG ($K_{rms}$ = 0.848). The optical seed pulses provided by the Ti:Sapphire laser had peak powers of up to about 10 kW with a pulse duration of 6 psec, which was wider than the electron bunch duration. Indeed, this ensures that the electron beam experiences a relatively uniform seed laser intensity over the entire bunch at the outset.

| **Electron Beam** | |
|---|---|
| Energy | 100.86 MeV |
| Bunch Charge | 360 pC |
| Bunch Duration | 1.8 psec |
| Peak Current | 300 A (parabolic pulse) |
| Emittance | 4 mm-mrad |
| rms Energy Spread | 0.1% |
| **NISUS Undulator** | weak focusing |
| Period | 3.89 cm |
| Amplitude (uniform) | 3.03 kG |
| $K_{rms}$ | 0.848 |
| Length | 10 m |
| Start Taper Point | 7.0 m |
| Optimal Taper | -4% |
| **Optical Field** | |
| Wavelength | 793.5 nm |
| Seed Power | 10 kW |
| Pulse Duration | 6 psec |

Table 3: Parameters for the BNL tapered wiggler experiment.

The experiment was run at the resonant energy. While the simulation can be run unambiguously at the resonant energy, finding the resonant energy in the experiment involved adjusting the electron beam energy from the linac. Since there was insufficient bunch charge to reach saturation in SASE mode, and since the growth rate peaks on-resonance, the unsaturated SASE interaction will yield maximum power when the beam energy is tuned to the resonant energy. As a result, when this condition was realized, the linac was "locked down" to this beam energy and the seed laser was turned on.

The pulse energy was measured by "kicking" the beam to the wall at various axial positions and measuring the output pulse energy that resulted. A comparison between the simulation and measured pulse energies for a uniform undulator is shown in Fig. 15, where the data (courtesy of X.J. Wang and J.B. Murphy) is indicated in red and the error bars indicate the standard deviation for a series of shots. It is evident that good agreement is found between the simulation and the measurements. Saturation at about 113 ± 28 μJ is found after about 7.0 – 7.5 m. The simulation result of 103 μJ is well within the range of uncertainty found in the experiment.

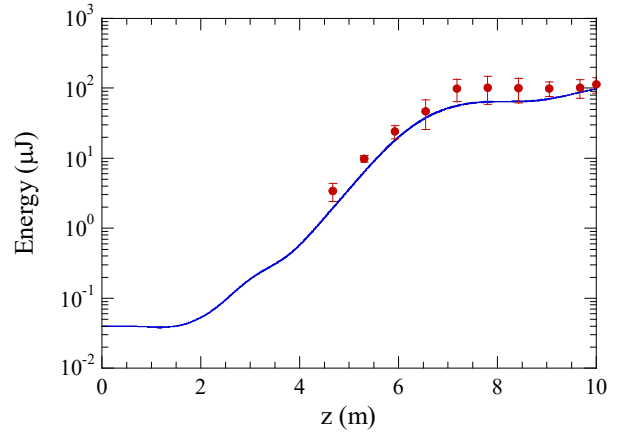

Fig. 15: Comparison between simulation and measured pulse energies for a uniform undulator (data courtesy of X.J. Wang and J.B. Murphy).

The NISUS undulator can be tapered in 1 meter steps. Since the optimal taper is dependent upon both the start-taper point and the taper slope, and since the start-taper point must be located prior to saturation in the uniform undulator, finding the optimal taper configuration was an iterative process. The choice of 10 kW seed power was made by trial and error to optimize the start-taper point at 7.0 m. Further optimization indicated that a down taper of 4% over the final 3 meters of the undulator yielded the maximum output power.

A comparison between the measured pulse energies for the tapered undulator (data courtesy of X.J. Wang and J.B. Murphy) and the corresponding simulation results is shown in red in Fig. 16. The uniform undulator results taken from Fig. 15 are also shown for comparison in blue. As evidenced in the figure, the agreement between the simulation and the measurements is excellent. The measured output was 283 ± 68 μJ, and the simulation result of 296 μJ also falls well within the range of experimental uncertainty. This represents an increase of almost 300% over the output of the uniform undulator.

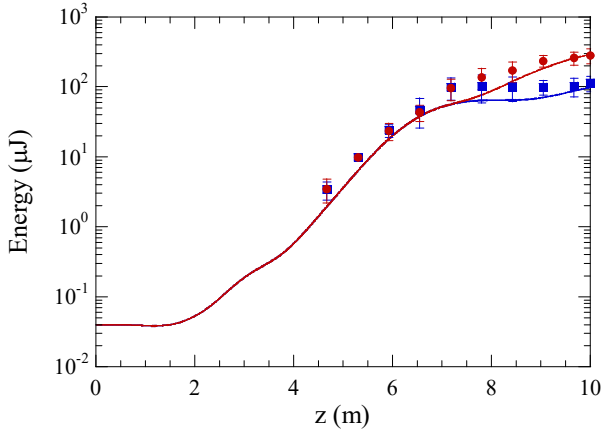

Fig. 16: Comparison between measured pulse energies and simulation results for the uniform (blue) and tapered (red) undulators (data courtesy X.J. Wang and J.B. Murphy).

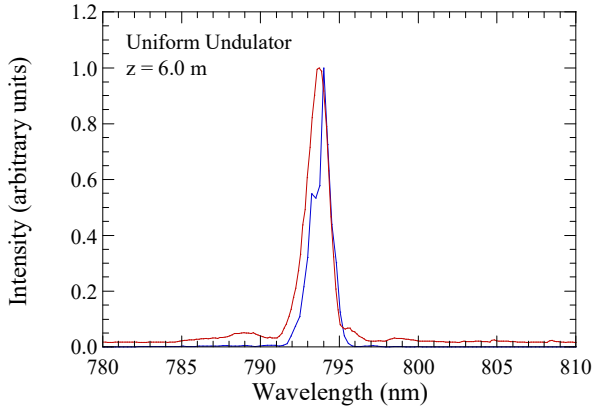

Fig.17: Measured (red) and simulated (blue) spectra in the uniform undulator section (data courtesy of X.J. Wang and J.B. Murphy).

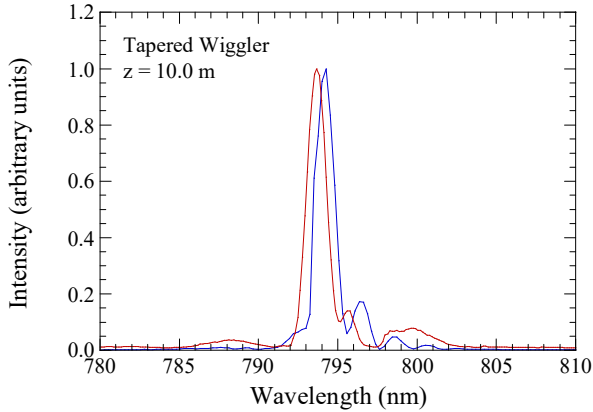

Fig. 18: Measured (red) and simulated spectra (blue) at the end of the tapered undulator (data courtesy of X.J. Wang and J.B. Murphy).

The spectra as observed in the experiment and as found in simulation show similar agreement. Note that the accuracy of the spectral measurements was limited by the bandwidth of the filter, and the experimental spectra may be shifted by as much as ± 0.5 nm. The spectra as determined at $z$ = 6.0 m are shown in Fig. 17 where the measured spectrum is shown in red and the simulation result is shown in blue. Observe that the peaks and spectral widths agree closely. The comparison between the observed (red) and simulation (blue) spectra at the exit from the tapered undulator is shown in Fig. 18, and the agreement is very good at this point as well. The shift in the simulation spectrum relative to that measured is only 0.45 nm, which is well within the sensitivity of the spectrometer. In addition, the spectral widths are very close, and the sidebands indicated at about 796 nm are also in good agreement between the measurement and the simulation. Although the simulation observed somewhat more sideband growth than the experiment, sidebands do not seem to be an important component of the output spectra.

## X. THE JLAB IR-UPGRADE FEL OSCILLATOR

To further investigate the simulation capabilities we also compared the simulation with the IR-Upgrade FEL oscillator at JLab [15]. The basic experimental parameters were a kinetic energy of 115 MeV, an energy spread of 0.3%, a bunch charge of 115 pC, a pulse length of 390 fsec, a normalized emittance of 9 mm-mrad in the wiggle plane and 7 mm-mrad in the plane orthogonal to the wiggle plane, and a repetition rate of 74.85 MHz for the electron beam. The planar undulator was 30 periods long, had a period of 5.5 cm, and a peak on-axis magnetic field of 3.75 kG. For proper electron beam transport through the undulator, we used a one period up- and down-taper. The electron beam was focused into the undulator with the focus at the center of the device. The resonator length was about 32 m and the cold-cavity Rayleigh length was 0.75 m. The total loss in the resonator was 21% with about 18% out-coupled per pass from the downstream mirror. For these settings, the wavelength was 1.6 μm.

To simulate the FEL oscillator, OPC takes the optical pulse at the exit of the undulator and propagates the pulse through the resonator and back to the entrance of the undulator. The FEL simulation takes this optical pulse and propagates it together with a fresh electron bunch through the undulator. This process repeats for a predefined number of roundtrips.

The length of the optical cavity must be selected so that the returning optical pulse is in synchronism with the electron bunches. The roundtrip time for the optical pulses in the cavity is $t_{roundtrip} = 2L_{cav}/c$ and the separation between electron bunches is $t_{sep} = 1/f_{rep}$, where $L_{cav}$ is the cavity length and $f_{rep}$ is the electron bunch repetition rate. Perfect synchronism is (referred to as zero-detuning) is obtained when $t_{roundtrip} = Mt_{sep}$, where $M$ is the number of optical pulses in the cavity. In this case there were 16 optical pulses in the cavity and the zero-detuning length is $L_0$ =

32.041946079 m. The cavity detuning curve is shown in Fig. 19 as a function of the difference between the cavity length L$_{cav}$ and the zero-detuning length. We find that the maximum output power of 14.52 kW occurs for a positive detuning of 2 µm and is close to the measured value of 14.3 ± 0.72 kW [15]. As a result, the predicted extraction efficiency is about 1.4%, which is close to the theoretical value of $1/2N_u \approx 1.7\%$. We remark that previous simulation of this experiment [24] yielded an average output power of 12.3 kW, and the present formulation is in better agreement with the experiment than in the earlier simulation. As in the previous simulation [24], the roughly triangular shape of the detuning curve is also in agreement with the experimental observation.

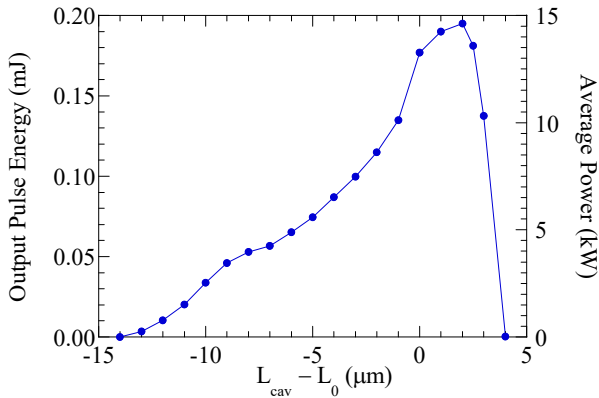

Fig. 19: The cavity detuning curve.

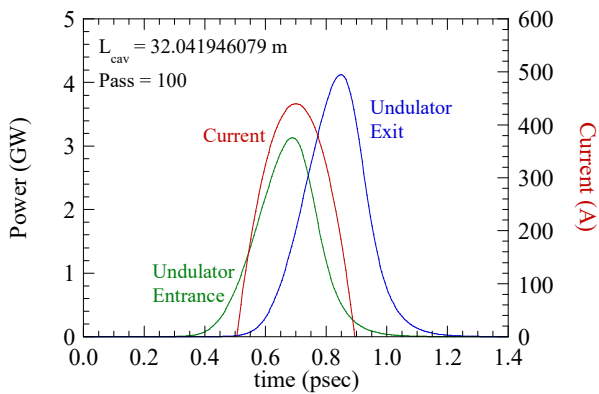

Fig. 20: Temporal profiles of the power in the optical pulse at the undulator entrance (green) and exit (blue) as well as the current in the electron bunch (right axis, red).

The temporal profiles of the optical pulse at the undulator entrance and exit as well as that of the electron bunch current are shown in Fig. 20 for the zero-detuning cavity length after pass 100 which corresponds to a stable, saturated steady-state. Observe that the electron bunch is centered in the time window, which has a duration of 1.4 psec. That this is at zero-detuning is indicated by the fact that the incoming optical pulse at the undulator entrance is in close synchronism with the electron bunch. It is also evident that the center of the optical pulse advances by about 0.16 psec as it propagates through the undulator, and this is in good agreement with the theoretical slippage estimate of $N_w\lambda/c$, where $N_w$ is the number of periods in the undulator. Finally, it should be remarked that this is in the steady-state regime where the losses in the resonator and the out-coupling are compensated for by the gain in the undulator.

## XI. SUMMARY AND CONCLUSION

In this paper, we have described a three-dimensional, time-dependent nonlinear formulation for the simulation of a variety of FEL configurations utilizing helical, planar, and elliptical undulators. For future reference, we refer to this formulation and simulation code as MINERVA. Along with the well-known three-dimensional representations of helical and planar undulators, a three-dimensional model is described to simulate an APPLE-II undulator. The simulation is in substantial agreement with a generalized parameterization for elliptical undulators. Comparisons of the simulation with a seeded infrared FEL amplifier, an infrared FEL oscillator, and SASE FELs operating at optical and x-ray wavelengths all showed good agreement with the experiments. Consequently, we feel that the formulation captures the basic physics of the FEL interaction over a wide range of parameters and can accurately, and with confidence, predict the performance of a large variety of FELs.

The Gaussian optical modes are not the ideal electromagnetic representation for all FELs. There is also interest in the development of FELs at spectral ranges that approach mm wavelengths. At wavelengths longer than 100 µm or so, the boundary conditions imposed by the walls of the drift tube cannot be satisfied using the Gaussian optical modes. Instead, a waveguide mode decomposition is more appropriate. Future development of this formulation will include a waveguide mode decomposition in addition to the Gaussian optical modes. Indeed, it is also intended to include a mixed decomposition appropriate where the waveguide boundary conditions are appropriate in one direction while the free-space modes are appropriate in the other direction. This will permit the simulation of long wavelength THz FELs using a rectangular drift tube which is compressed in one direction but relatively open in the other. These developments will be reported in future publications.

## ACKNOWLEDGEMENTS

The authors acknowledge helpful discussions with S.V. Benson, D.J. Dunning, P. Emma, L. Giannessi, J.B. Murphy, H.-D. Nuhn, D. Ratner, M. Shinn, N.R. Thompson, and X.J. Wang.